\documentstyle[editedvolume,numreferences]{crckapb} 
\input{psfig}

\newcommand{\skyview}{{\sf SkyView\ }}
\newcommand{\aladin}{{\sf Aladin\ }}

\begin{opening}
\title{CROSS WAVELENGTH COMPARISON OF IMAGES AND CATALOGS}

\author{J.G. Bartlett}
\author{D. Egret}
\institute{Centre de Donn\'ees astronomiques de Strasbourg\\
           11, rue de l'Universit\'e, 67000 Strasbourg, FRANCE\\
	   bartlett@astro.u-strasbg.fr; egret@astro.u-strasbg.fr}

\end{opening}

\runningtitle{CROSS WAVELENGTH COMPARISON OF IMAGES AND CATALOGS}

\begin{document}


\section{Introduction}

	We would like to start this discussion by attempting to
make a useful, working distinction between {\it catalogs} and
{\it databases} of astronomical objects.  It seems to us
that such a distinction could be made based upon the mode
of access to the information:  A catalog may be considered 
as a {\it list} of objects, almost invariably ordered by 
coordinates; a database, on the other hand, may be distinguished
by its ability to extract a set of objects based on user-given
criteria, such as all objects within a certain sky region with
magnitudes brighter than $m$ in the blue.  Catalogs form the basis
of the database, which adds the means of multi-criteria access to
the information.  In concrete terms,
one usually thinks of SIMBAD, NED and LEDA as databases, while
an ftp site containing electronic lists of objects may be
thought of as a catalog storage warehouse.  

	The next comment we would like to make is related to the
above definition and concerns the sizes of astronomical catalogs.
In Table 1, we list the numbers of objects expected or currently
existing in various catalogs.  The list is by no means exhaustive, 
but only meant to give an idea of the number of cataloged objects as 
a function of spectral domain.  It is clear that there is a
large dichotomy between the Optical/Near-infrared (NIR) and the rest
of the spectrum: for the former, one expects on the order
of $10^8-10^9$ sources, while in no other part of the 
spectrum is there more than a {\it few} $10^6$ sources.
In terms of numbers and storage volume, {\it the Optical/NIR
dominates}, and this has important consequences for both 
the access to and the science performed with astronomical
catalogs.  

\begin{table}[htb]
\begin{center}
\caption{Number of Cataloged Objects by Spectral Domain}
\begin{tabular}{llll}
\hline
Optical/NIR       &                 & Other  &         \\
\hline
GSC-II:           & $10^9$          & FIRST: & $10^6$  \\
USNO A1:          & $5\times 10^8$  & IRAS:  & $10^5$  \\
SuperCOSMOS/APM:  & $5\times 10^8$  & ROSAT: & $10^5$  \\
DENIS/2MASS:      & $10^9$          &        &         \\
\hline
\ & $\sim 10^8$ - $10^9$            &        & $<10^6$ \\
\end{tabular}
\end{center}
\end{table}

	What is the origin of this dominance?   As an answer to 
this question, we will take the opportunity to defend a little
the oft-attacked photographic plate.  The advantages of CCDs 
should not obscure the fact that the photographic plate 
has proven itself as a magnificent detector, combining a large 
field-of-view with high spatial resolution; {\it and} it is its 
own storage medium.  Remember that a $6^o\times 6^o$ Schmidt 
plate with 1 arcsec resolution represents $4.7\times10^8$ pixels 
taken in a single exposure!  No other detector to date can 
cover the focal plane with more efficiency.  Although modern 
detectors are beginning to achieve the same sky coverage
and resolution as photographic plates by scanning CCD arrays 
across the sky (such as DENIS, 2MASS and the Sloan Digital 
Sky Survey), it seems to us arguable that part of the 
reason for the dominance of the Optical/NIR rests with the
one-hundred year legacy of photographic plates.
  
	Apart from the serious, it must be said, drawbacks of 
nonlinearity and calibration difficulties, photographic plates 
present a problem of access: the information is not readily 
manipulated, as is a computerized image.  This has changed 
with the advent of plate scanning machines, which produce the 
desired electronic quantification.  However, this digitization
program suffers from the need of adequate storage media
for the terabytes of data that a full sky Optical/NIR survey
produces; this problem highlights the performance of a 
photographic plate as a storage medium, the equal of which,
in terms of the ratio (viability)/(volume) over long periods of time,
is as yet unmatched.  

	In fact, the key challenge facing astronomical archiving
at present is this question of storage and access to such large
quantities of data.  As an example, the ``classical'', if such
a term is now permitted, databases such as SIMBAD, NED or LEDA
contain at most a few $\sim 10^6$ objects; multi-criteria access
to a much larger number of objects seems to demand novel 
techniques, and forms a large part of the development effort
of the Sloan Survey (Szalay, Brunner - these proceedings).  
An interesting question
in this context is the relation between efficient storage
and science.  Designing an archive structure for quick access
requires some knowledge of the correlations and clustering
of the data in the space defined by the measured parameters.
This amounts to a statistical analysis of the survey objects'
properties.

	Let us finish this section by returning to the implications
of the Optical/NIR dominance.  One important consequence of this 
dominance is that it would seem to guarantee that 
most cross-identification efforts of newly detected objects
in other wavebands will rely Optical/NIR catalogs.  The questions
we would pose are:  ``Is it useful to have a centralized {\it database}
of the ensemble of Optical/NIR objects?''; and, if so, ``Who
has the resources to construct it?''  These questions concern
the construction of a true multi-criteria {\it database},
the fundamental elements of which, as we have seen, 
require something beyond that presently used by NED/SIMBAD/LEDA.

\section{Cross Identifications - Nature of the Problem}

	In this section, we would like to discuss some statistical
aspects of the general problem of identifying objects 
from two different catalogs, a procedure known as the cross--identification
of sources.  The two catalogs will be called {\it Letters} and
{\it Numbers}.

\subsection{As a Test of Hypotheses}

	Consider the situation depicted in Figure 1, where 
we wish to cross-identify a source, $A$, from catalog {\it Letters}
with objects from the catalog {\it Numbers}, two of which are shown 
within the error ellipse of source A (sources 1 and 2).  
Key to the problem are accurate positions in the two catalogs, because
spatial proximity serves as a primary criterion.  After 
selecting objects for consideration based on coordinates,
one may think to apply supplementary criteria; for example,
suppose that $A$ were an IRAS source with
galaxy colors, and that source 1 is an optical galaxy,
while source 2 is a star.  This additional information
would favor an identification with the optical galaxy.

\begin{figure}
\centerline{\psfig{figure=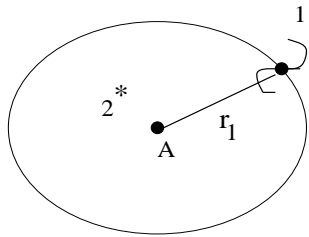,width=4cm,height=3cm}}
\caption{Geometry for the cross-identification of source A
	 with source 1.}
\end{figure}

	A quantitative approach must rely on a measure of the
acceptability of a given cross-identification.  If we ignore,
for simplicity, any supplemental information, we may define
the {\it likelihood} that source 1 is the same as source A
as the {\it probability} that the two cataloged positions would
be separated by distance $r_1$ {\it if they represented 
the same physical object}.  Assume that the position of 
source 1 (and 2) is much more precise than that of
source A; and, further, that the error in position A is
described by a two-dimensional Gaussian.  Then,
\begin{equation}
\label{H1}
L_1 \equiv \bar{N}(<m_2) e^{-\bar{N}(<m_2)}
           \frac{1}{2\pi |M|^{1/2}}
	e^{-\frac{1}{2}\vec{r_1}^T\cdot M^{-1} \cdot \vec{r_1}},
\end{equation}
where $M$ is the covariance matrix and $|M|$ is its
determinant.  The Poisson term expresses the probability
of finding one source from {\it Numbers} by chance alignment
within $r_1$ at magnitudes brighter than $m_2$, when the
expected number is $\bar{N}(<m_2)$.  We denote this 
likelihood by the subscript $1$, which will hereafter
refer to the {\it hypothesis} $H1$ that sources 1 and A are
the same.   Notice that this procedure requires a clear 
quantitative statement of the positional errors ({\it
i.e.}, a Gaussian distribution).  
The alternate hypothesis, that sources 1 and A are
separated by distance $r_1$ by chance alignment, may be assigned
the likelihood
\begin{equation}
\label{H0}
L_0 \equiv \bar{N}(<m_2) e^{-\bar{N}(<m_2)} n(m_1).
\end{equation}
Here, $n(m_1)$ denotes the density of objects in the reference
catalog ({\it Numbers}) at the magnitude of source 1, $m_1$.
The subscript $0$ will denote the hypothesis $H0$ that
the sources, 1 and $A$, are not the same physical object.  

\begin{figure}
\centerline{\psfig{figure=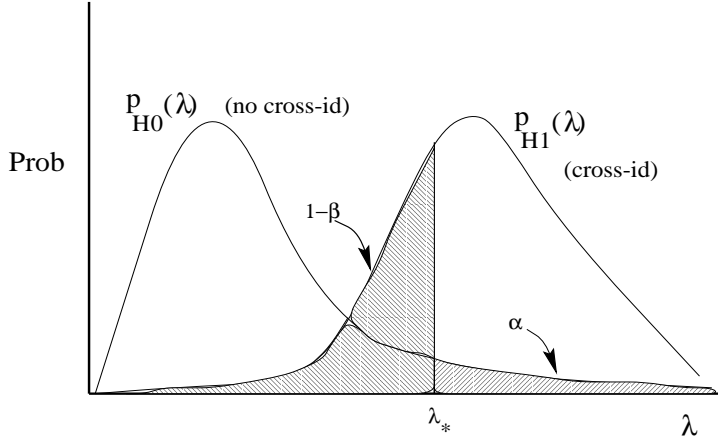,width=10cm,height=6.2cm,bbllx=0pt,bblly=0pt,bburx=551pt,bbury=360pt,clip=}}
\caption{Likelihood Ratio Test: The distributions $p_{H0}(\lambda)$
	 and $p_{H1}(\lambda)$ are shown and labeled (see text).  
	 The shaded area to the right of 
	 $\lambda_*$ represents $\alpha$ and that to the left
	 represents $1-\beta$.}
	
\end{figure}

	The decision to cross-identify sources 1 and A 
now amounts to the statistical rejection of the null 
hypothesis, $H0$.  For this purpose, we may employ
the {\it likelihood ratio test} (Meyer 1975), which focuses on the 
quantity $\lambda = L_1/L_0$.  We must now construct the
probability distributions of $\lambda$ under the assumptions 
$H0$ ($p_{H0}(\lambda)$) and $H1$ ($p_{H1}(\lambda)$).
This step {\it requires} a set, sometimes called 
the {\it training set}, of previously known, sure-fire 
cross-identifications.  We emphasize the importance
of this training set and the fact that the need for
such {\it a priori} information is unavoidable in 
any quantified approach.  Suppose, then, that we have found 
the two distributions and that they resemble the curves 
shown in Figure 2.  

	We identify source A with source 1 
with {\it confidence} $\alpha$ if the observed value 
$\lambda_{obs} > \lambda_*$, where $\lambda_*$ is defined by 
$P_{H0}(\lambda>\lambda_*)=\alpha$ (note that capital $P$ 
refers to the cumulative distributions).  The 
confidence $\alpha$ represents the probability of a 
type I statistical error, or the chance that the cross--identification
is wrong.  The {\it power}, $\beta$, of 
our cross-identification is defined by 
$\beta = P_{H1}(\lambda>\lambda_*)$; the quantity
$1-\beta$ represents the probability of a type 
II statistical error and embodies the concept of 
the {\it reliability} of a conclusion that
the sources are {\it not} the same.  A wonderful
example of this kind of approach is given by 
Lonsdale {\it et al.} (1996), and we are grateful to 
C. Lonsdale for helpful discussion on this topic.

\subsection{As a Maximization of Likelihood}

	In the case just described, the cross-identification
was made without regard to the outcome of attempts to 
cross-identify other sources.  This
approach might be readily applied when catalog {\it Letters}
contains many fewer objects than catalog {\it Numbers}
(a situation we may describe by saying that ``catalog
{\it Numbers} is dense within catalog {\it Letters}'').
Consider another case in which the two catalogs each have 
about the same source density, as schematically represented
in Figure 3.  In pursuing the cross-identification of source
A with source 1, it now seems particularly urgent to incorporate
the possibility that source 1 could also be identified with source B.
One way of proceeding is to calculate the probability,
$P_i$, $i=1,7$, of each of the seven possible outcomes of
the cross-identification of the two catalogs: $(A,1)(B,2)$,
$(A,1)(B,0)$, $(A,2)(B,1)$, $(A,2)(B,0)$, $(A,0)(B,1)$,
$(A,0)(B,2)$, $(A,0)(B,0)$, where a $0$ means that the source 
in {\it Letters} has no counterpart in catalog {\it Numbers}.  
The case $i$ which has the largest probability represents the
{\it most likely} cross-identification of the two catalogs,
and therefore the proper choice.  This approach is global 
in that it seeks the most satisfactory solution for all 
the sources considered at once.  
Note that this method is well adapted to treat systematic
problems, such as a coordinate offset between the two
catalogs; in fact, it incorporates what we mentally
do to align two sky charts.  The drawback of this approach,
as formulated, is the large computing power required to 
consider all possible cross--identifications.

\begin{figure}
\centerline{\psfig{figure=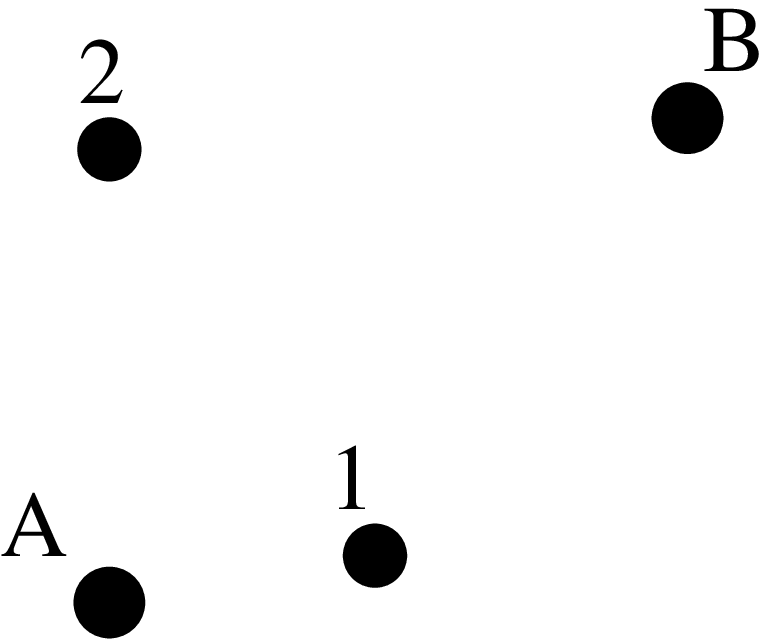,width=5cm,height=3cm}}
\caption{}
\end{figure}

\section{Two Software Systems}

	In this section, we present the characteristics of 
two software systems designed to aid the astronomer 
in making cross-identifications.  These packages are interactive,
and therefore not suitable for performing automatic 
cross-identifications of large catalogs.  
Their philosophy is, rather, to replace by interactive
software the tasks of going to the library to obtain
and combine copies of images ({\it e.g.}, 
sky survey films), tables and catalogs for a specific 
region of the sky.  The second package, \aladin, is 
particularly well adapted for the training/optimization 
of automated routines (recall the importance of the
training set in Section 2), the cross-identifications of small,
user catalogs and the resolution trouble cases, which is
useful for astronomical database quality control and for
``mopping-up'' difficult objects flagged by an automatic
routine.   

\subsection{SkyView - {\tt http://skyview.gsfc.nasa.gov/skyview.html}}

	This package is provided by the NASA Goddard Space Flight
Center (McGlynn {\it et al.} 1996).  \skyview exists in
true interactive form or as a Web tool.  On the Web, the
user fills out a form requesting images and catalogs of 
interest; \skyview then returns a composite image overlaying
the requested images and the positions of the cataloged
objects in the region.  Available images cover the 
entire electromagnetic spectrum, from Gamma rays (EGRET) to
the radio, and may be superimposed by adjusting a color table
or by plotting the contours of one image on another.  A 
large range of catalogs are on-line.  In addition 
to plotting the positions of the selected catalog objects,
\skyview returns a table containing the stored information on 
each object.  Access to SIMBAD as a name resolver
permits the user to select a region of sky by giving
the name of a well known object instead of sky coordinates.

\subsection{Aladin - {\tt http://cdsweb.u-strasbg.fr/CDS.html}}

	{\sf Aladin} is an interactive sky atlas under development
at the {\it Centre de Donn\'ees astronomiques de Strasbourg} 
(CDS) (Bonnarel {\it et al.} 1996).  \aladin provides
unified access to the CDS archives by overlaying the positions
of cataloged objects on digitized images of Schmidt plates.
To obtain the information, the \aladin client queries the CDS 
SIMBAD, catalog ($\sim 1500$ catalogs) and image archive servers.
Additionally, the client accepts both user defined catalogs, 
permitting the astronomer to visualize his sources in 
conjunction with documented information, and user given
image files, provided they conform to the World Coordinate System 
FITS extension. The image archive at CDS consists of 
the Space Telescope Science 
Institute's {\it Digital Sky Survey}, first and second epochs
(as this becomes available), and higher resolution scans
(0.7 arcsecs/pixel) of the Southern Galactic Plane, Ecliptic 
Poles and the Magellanic Clouds; these latter images are provided
by the MAMA facility at the Paris Observatory and by 
SuperCOSMOS of the Royal Observatory, Edinburgh.  All of the
images include astrometric {\it and} photometric calibrations.

	Beyond its capability of visualization, \aladin
furnishes an active approach to cross-identification:
The system includes image analysis tools for the detection, 
extraction and classification of objects found on the image.
For example, the user may plot the positions of his sources and
then request that all objects on the image found within a specified
aperture, centered on his positions, be extracted and associated 
with the corresponding sources in a table, which may then
be written to disk.  In this way, he directly creates a new 
catalog of his sources cross-identified with objects from a 
pre-selected catalog.

\section{Acknowledgments}

	We are very grateful for the helpful discussions that we 
enjoyed with F. Bonnarel, C. Lonsdale and S. Mei.

\end{document}